\documentclass[a4paper,11pt]{article}
\usepackage{pos}
\definecolor{YellowOrange}{rgb}{1.0, 0.72, 0}
\definecolor{MyDarkGreen}{HTML}{006766}
\newcommand{\red}[1]{{\textcolor{red}{#1}}}

\makeatletter
     {\bgroup\raggedright\small\section*{\refname
        \@mkboth{\MakeUppercase\refname}{\MakeUppercase\refname}}%
      \list{\name{bib\@arabic\c@enumiv}
            \@biblabel{\@arabic\c@enumiv}}%
           {\settowidth\labelwidth{\@biblabel{#1}}%
            \setlength\itemsep{0pt}
            \setlength\parskip{0pt}
            \setlength\parsep{0pt}
            \setlength\partopsep{0pt}
            \leftmargin\labelwidth
            \advance\leftmargin\labelsep
            \@openbib@code
            \usecounter{enumiv}%
            \let\p@enumiv\@empty
            \renewcommand\theenumiv{\@arabic\c@enumiv}}%
      \sloppy\clubpenalty4000\widowpenalty4000%
      \sfcode`\.\@m}
     {\def\@noitemerr
       {\@latex@warning{Empty `thebibliography' environment}}%
      \endlist\egroup}
\makeatother

\title{Symmetric Mass Generation}

\author[a]{Anna Hasenfratz}
\author*[b]{Oliver Witzel}
\onbehalf{\newline Lattice Strong Dynamics collaboration}
\affiliation[a]{Department of Physics, University of Colorado, 
Boulder, Colorado 80309, USA}

\affiliation[b]{Theoretische Physik 1, Center for Particle Physics Siegen, Naturwissenschaftlich-Technische Fakult\"at, Universit\"at Siegen, 57068 Siegen, Germany}

\emailAdd{anna.hasenfratz@colorado.edu}
\emailAdd{oliver.witzel@uni-siegen.de}

\abstract{In recent years tantalizing signs for a novel phase have been reported that is chirally symmetric but nevertheless exhibits massive bound states. The necessary condition for such a phase, referred to as Symmetric Mass Generation (SMG), is the cancellation of all (continuous and discrete) 't~Hooft anomalies. In 3+1 dimensions this occurs in systems containing a multiple of 16 massless Weyl fermions. SMG was originally discovered in lower dimensional condensed matter systems. 
We present results investigating four dimensional field theories with gauge group SU(3). Our findings suggest that SU(3) with $N_f=8$ fundamental fermions exhibits an SMG phase not only on the lattice but also in the infinite cutoff continuum limit. If confirmed, SMG could provide a new UV completion of the standard model and give rise to new scenarios for beyond standard model physics.}

\FullConference{The European Physical Society Conference on High Energy Physics (EPS-HEP2025)\\
7-11 July 2025\\
Marseille, France\\}

\begin{document}
\maketitle
\section{Introduction}

In QCD, hadron masses arise from the dynamical breaking of chiral symmetry. For $N_f$ massless quark flavors, the $SU(N_f)_L \times SU(N_f)_R$ symmetry is spontaneously broken by the quark condensate $\langle \bar{\psi}\psi\rangle \neq 0$ to the vector subgroup $SU(N_f)_V$. As a result, there are $N_f^2-1$ massless Nambu–Goldstone bosons --- the pions --- while all other hadrons acquire masses proportional to the condensate scale.

In recent years, an alternative mechanism known as \emph{symmetric mass generation} (SMG) has been identified~\cite{Fidkowski:2009dba} (for a recent review, see~\cite{Wang:2022ucy}). In an SMG phase, bound states become massive without forming a bilinear condensate. Instead, strong four-fermion or gauge–Yukawa interactions generate a four-fermion condensate that gaps the system while preserving chiral symmetry. SMG can occur only for symmetries free of ’t~Hooft anomalies. In most known examples, the axial $U(1)_A$ symmetry is explicitly broken to $\mathrm{Spin}\text{-}Z_4$ by four-fermion interactions, and the remaining anomaly in four dimensions is canceled when the number of Dirac fermions in the ultraviolet (UV)  is a multiple of eight~\cite{You:2014oaa,Garcia-Etxebarria:2018ajm}. Thus, for fundamental fermions the necessary condition for an SMG phase is
$N_f \times N_c \;\text{mod}\; 8 = 0$.

\begin{figure}[!htb]
\centering
\begin{picture}(140,40)(0,-4)
  \put(9,30){\textcolor{YellowOrange}{strong coupling}}
  \put(5,25){\linethickness{3mm}\textcolor{YellowOrange}{\line(1,0){60}}}
  \put(5,20){\textcolor{YellowOrange}{$\blacktriangleright\;$} SMG phase}
  \put(5,15){\textcolor{YellowOrange}{$\blacktriangleright\;$} Chirally symmetric} 
  \put(5,10){\textcolor{YellowOrange}{$\blacktriangleright\;$} Confining, gapped spectrum}
  \put(67.5,25){\linethickness{3mm}\red{\circle*{4}}}
  \put(58,20){\red{continuous}}
  \put(55,15){\red{phase transition}}
  \put(67.5,14){\thicklines\red{\vector(0,-1){8}}}
  \put(52,3){continuum limit exist}
  \put(58,-1){RG $\beta$ function}  
  \put(70,25){\linethickness{3mm}\textcolor{MyDarkGreen}{\line(1,0){60}}}
  \put(77,30){\textcolor{MyDarkGreen}{weak coupling}}
  \put(85,20){\textcolor{MyDarkGreen}{$\blacktriangleright\;$} Appears conformal}
  \put(85,15){\textcolor{MyDarkGreen}{$\blacktriangleright\;$} Chirally symmetric}
  \put(85,10){\textcolor{MyDarkGreen}{$\blacktriangleright\;$} Conformal hyperscaling}  
\end{picture}
\caption{Sketch of the phase structure for SU(3) with $N_f=8$ fundamental flavors. A similar phase structure was identified for SU(2) gauge with $N_f=4$ in Ref.~\cite{Butt:2024kxi}.}
\label{Fig.Sketch}
\end{figure}
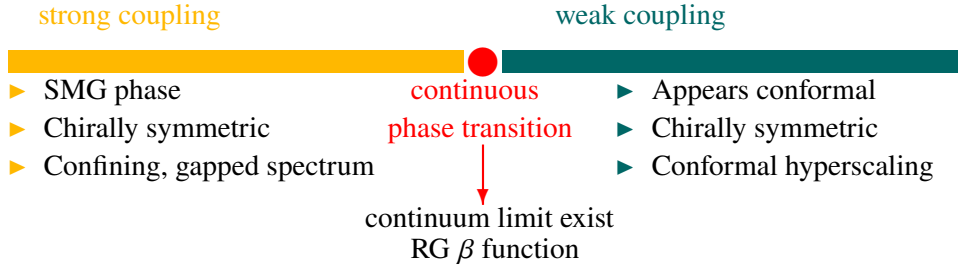

The simplest gauge–fermion systems where an SMG phase may appear are $SU(2)$ gauge theory with $N_f=4$ and $SU(3)$ gauge theory with $N_f=8$ fundamental fermions. Numerical simulations with staggered lattice fermions show remarkably similar phase diagrams for these two systems, as sketched in Fig.~\ref{Fig.Sketch}~\cite{Butt:2024kxi,Hasenfratz:2024fad}. At weak coupling lies an apparently conformal phase (green) exhibiting hyperscaling and unbroken chiral symmetry. It is separated by a continuous phase transition from a strong-coupling SMG phase (yellow), which is also chirally symmetric but gapped and confining. Finite-size scaling indicates that the transition is continuous, implying the existence of an ultraviolet fixed point (UVFP) where a continuum, strongly coupled renormalized QFT can be defined.

Here we focus on the $SU(3)$ gauge theory with $N_f=8$ fundamental flavors. Our simulations in the chiral limit show no evidence of chiral symmetry breaking, even on the largest volumes studied. This challenges the long-held view that the eight-flavor system lies just below the conformal sill and could serve as a walking composite Higgs candidate. 

The $SU(3)$, $N_f=8$ system has been extensively investigated in large-scale numerical studies~\cite{LatticeStrongDynamics:2020uwo,LatticeStrongDynamics:2023bqp,LatKMI:2016xxi,LatKMI:2025kti}. These simulations, performed at finite fermion mass, aimed to distinguish explicit from spontaneous chiral symmetry breaking.
In contrast, our calculations are carried out directly in the chiral limit, and incorporates additional heavy Pauli–Villars fields~\cite{Hasenfratz:2021zsl}, which suppress gauge-field fluctuations. This improvement of the lattice action allows us to reach substantially stronger renormalized couplings than in previous work. For illustration, we compare the finite-volume gradient flow (GF) coupling $g^2_{GF}(\tau,L)$~\cite{Luscher:2010iy,Fodor:2012td} at flow time $\tau = (c\,L)^2/8$, $c=0.25$, on $L/a = 24$ lattices. In the chiral limit, the lattice action used in Refs.~\cite{LatticeStrongDynamics:2020uwo,LatticeStrongDynamics:2023bqp} yields $g^2_{GF}(L/a=24)\approx 17$ at bare coupling $\beta_b \equiv 6/g_0^2 = 4.8$; stronger couplings are inaccessible due to a first-order transition at $\beta_b \gtrsim 4.7$. In contrast, our PV-improved action exhibits a steadily increasing coupling up to $g^2_{GF}(L/a=24) \!\approx\! 26$, where a continuous transition to the SMG phase occurs. Throughout the weak-coupling region, the system displays conformal hyperscaling even on our largest volume, $32^3 \times 64$. The transition observed at $g^2_{GF}(L/a=24)\!\approx\!26$ is  beyond the reach of prior simulations unless the lattice size is increased by over an order of magnitude.

We emphasize that our numerical results are consistent with prior studies where there is an overlap, but lead to a different conclusion about the infrared dynamics, since we can probe much stronger couplings.

The simulations of Refs.~\cite{Butt:2024kxi,Hasenfratz:2024fad,LatticeStrongDynamics:2020uwo,LatticeStrongDynamics:2023bqp} employ one-component staggered lattice fermions. Because of fermion doubling, a single staggered field produces sixteen low-energy modes, which assemble into four Dirac spinors. The resulting action explicitly breaks the axial $U(1)_A$ and most of the flavor (taste) symmetries of the continuum theory but preserves a remnant $U(1)_\epsilon$ corresponding to a pseudoscalar–pseudoscalar (spin–taste) rotation. These symmetry-breaking operators are irrelevant near the Gaussian fixed point, so the continuum limit of one staggered field describes four Dirac fermions perturbatively. However, if any of these operators become relevant at a strongly coupled nonperturbative fixed point, the infrared dynamics can differ dramatically. In that case, the continuum limit may describe a novel strongly coupled QFT realizing the SMG phase with emergent symmetries distinct from those of the perturbative QFT. Detailed discussions of this mechanism and its implications are presented in Ref.~\cite{Cenke}.

\section{Numerical setup and results}

We use the Symanzik gauge action together with nHYP-smeared~\cite{Hasenfratz:2007rf} staggered fermions to generate dynamical gauge-field configurations with eight fundamental flavors. To enable simulations at very strong coupling ($g^2 \gtrsim 20$), we suppress gauge-field fluctuations by introducing additional Pauli–Villars fields~\cite{Hasenfratz:2021zsl}. All simulations are carried out in the chiral limit, with bare couplings $\beta_b = 6/g_0^2$ ranging from 9.50 (weakest coupling) to 8.30 (strongest coupling), covering both the weak-coupling conformal phase and the strong-coupling SMG phase. Simulations are performed on $(L/a)^3 \times (T/a)$ lattices, and we present results from $16^3 \times 32$, $24^3 \times 48$, and $32^3 \times 64$ volumes, while additional runs on $48^3 \times 96$ lattices are ongoing.

\begin{figure}[tbh]
    \centering
    \includegraphics[width=0.49\linewidth]{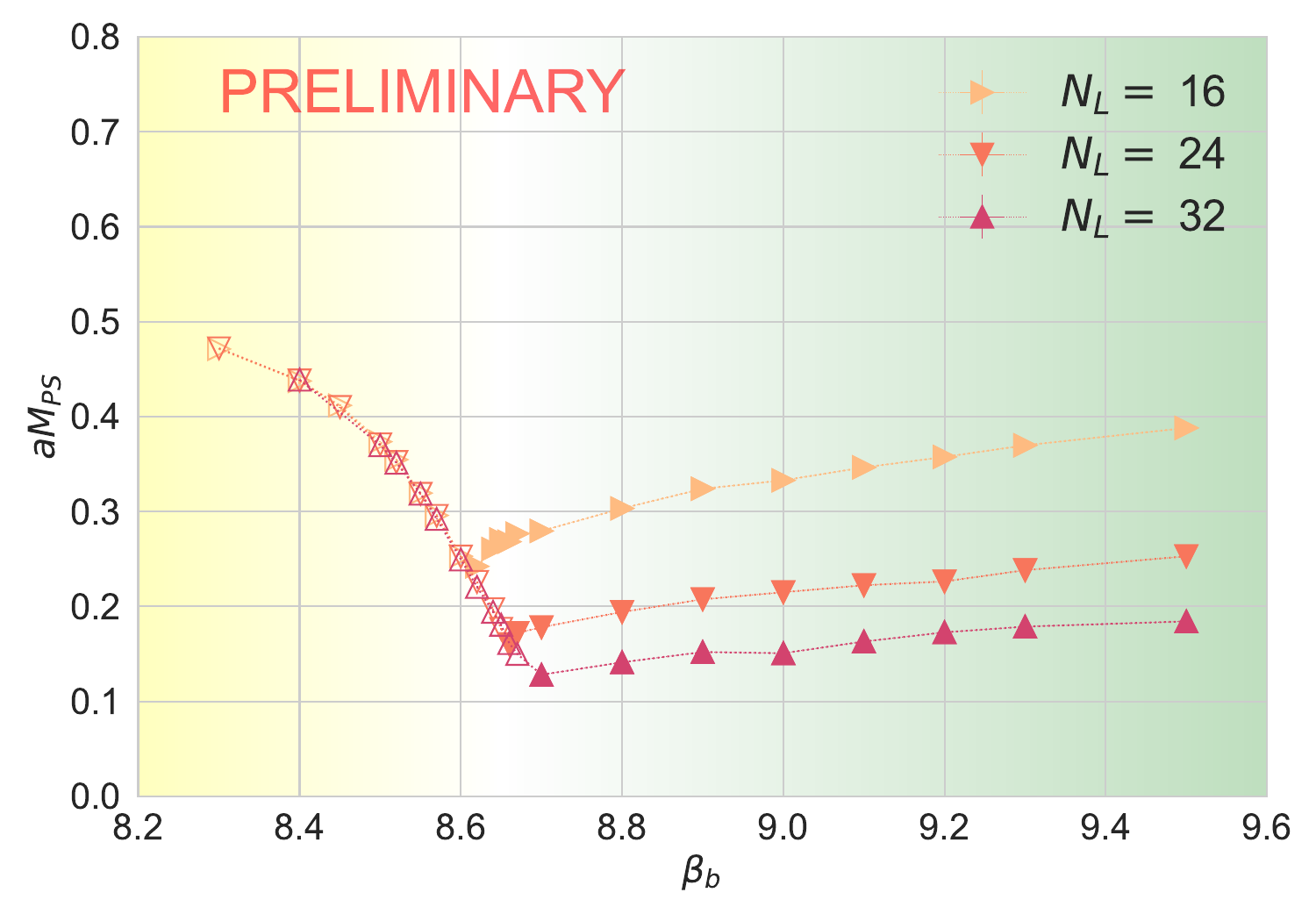}
     \includegraphics[width=0.49\linewidth]{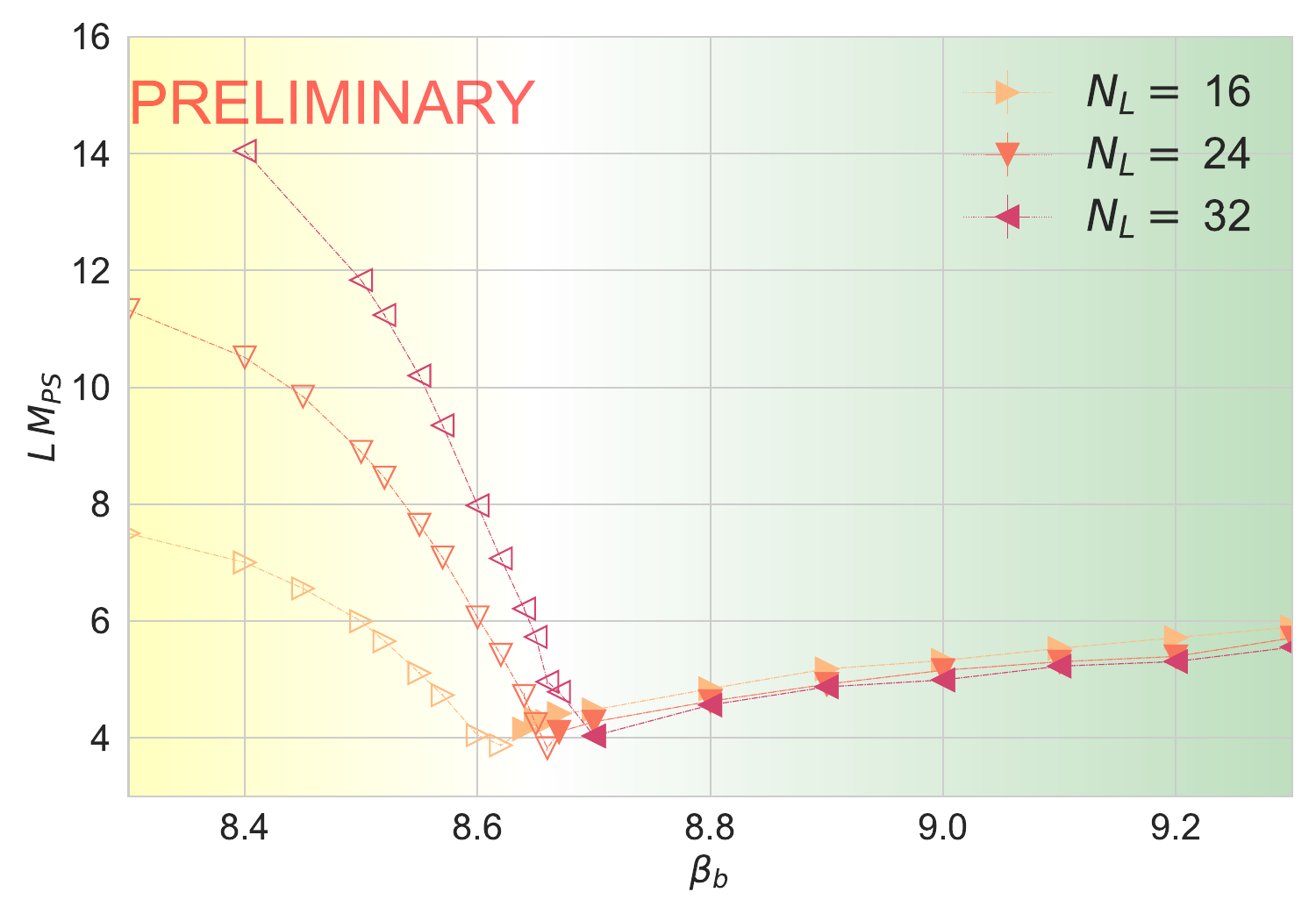}
    \caption{Left: lightest pseudoscalar mass $aM_{PS}$ in lattice units vs.~the bare gauge coupling $\beta_b$. Right: Test for conformal hyper-scaling by plotting $L\cdot M_{PS}$ vs.~$\beta_b$.}
    \label{Fig.M_PS-scaling}
\end{figure}

Our analysis begins with the low-lying meson spectrum, focusing on the lightest pseudoscalar state shown in Fig.~\ref{Fig.M_PS-scaling}. The left panel displays the pseudoscalar mass in lattice units versus the bare gauge coupling $\beta_b$. In the strong-coupling regime ($\beta_b \lesssim 8.7$), the results from all three volumes collapse onto a single curve, indicating that the pseudoscalar mass is volume independent and the system is gapped. In contrast, at weaker coupling ($\beta_b \gtrsim 8.7$), the pseudoscalar mass exhibits volume scaling consistent with conformal behavior. To test conformal hyperscaling, the right panel shows $L \cdot M_{PS}$: while the strong-coupling data separate by volume, the weak-coupling data collapse onto a single curve, confirming the expected scaling behavior in the conformal regime.

\begin{figure}[tb]
    \centering
    \includegraphics[width=0.49\linewidth]{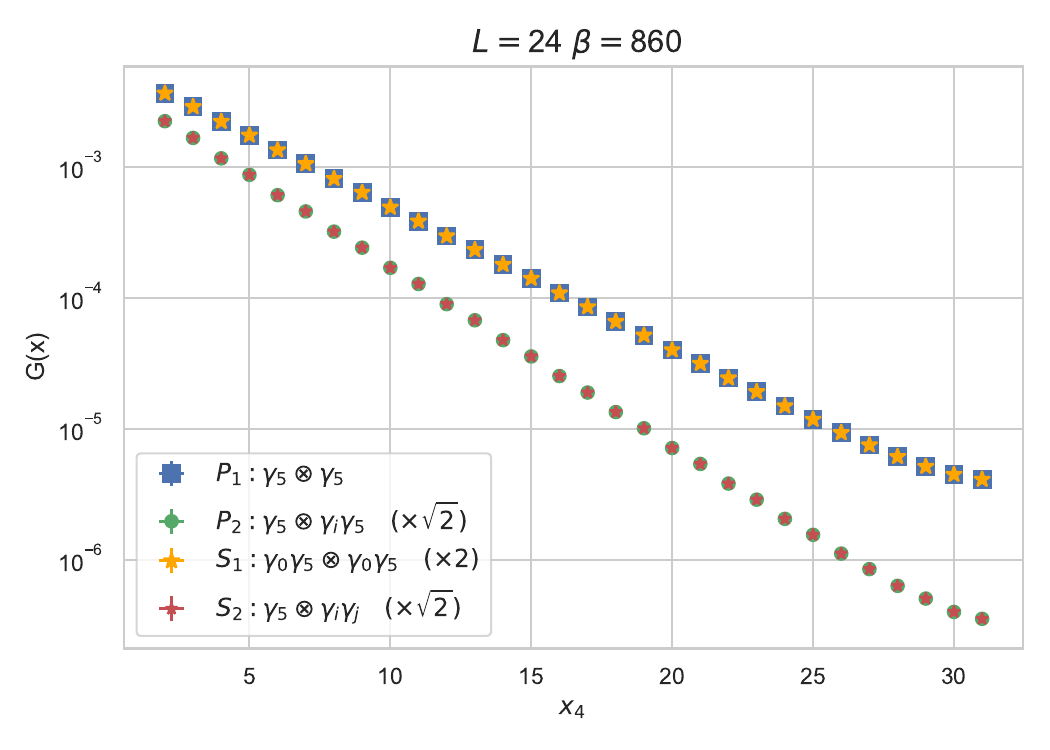}
     \includegraphics[width=0.49\linewidth]{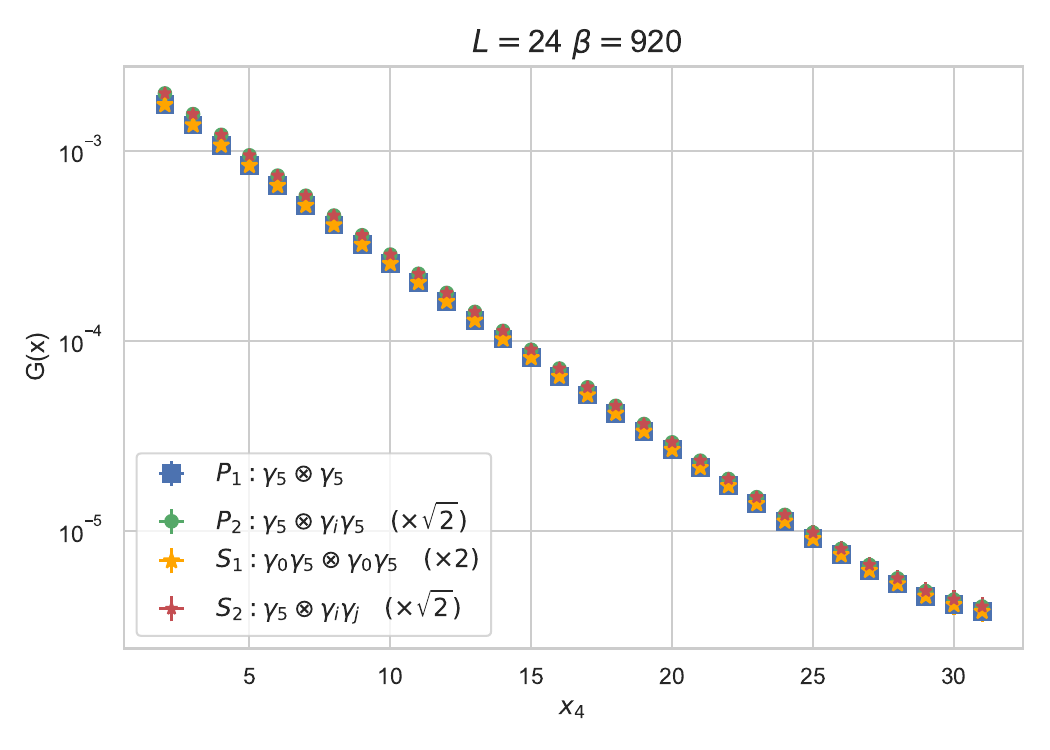}
    \caption{Pseudoscalar and scalar correlators on $24^3\times 48$ lattices at strong coupling ($\beta_b=8.60$) on the left and weak coupling ($\beta_b=9.20$) ont the right.}
    \label{Fig.M_PS-correlators}   
\end{figure}

Next, we examine the two-point correlation functions. Figure~\ref{Fig.M_PS-correlators} shows the pseudoscalar and scalar correlators as functions of the Euclidean time separation $x_4$, obtained on $24^3 \times 48$ lattices. The left panel corresponds to strong coupling ($\beta_b = 8.60$), while the right panel shows results at weak coupling ($\beta_b = 9.20$). In both cases, the pseudoscalar and scalar channels are degenerate, demonstrating exact parity doubling that persists to machine precision, configuration by configuration. At weak coupling, we  observe that taste splitting is strongly suppressed, while in the strong-coupling SMG phase it becomes significantly larger, suggesting enhanced taste breaking near the SMG fixed point. Similar behavior is observed for other channels, including the lightest vector and axial-vector mesons.

Finally, we investigate the nature of the phase transition. We restrict our analysis to the range $8.55 \lesssim \beta_b^{[24]} \lesssim 8.66$, near the critical point within the gapped SMG phase. By performing a finite-size scaling (FSS) curve-collapse fit, we rescale data from the $16^3 \times 32$ and $32^3 \times 64$ volumes to match the units of the $24^3 \times 48$ ensemble. FSS predicts that the renormalization-group–invariant quantity $L \cdot M(\beta_b; L)$ is described by a unique function of the scaling variable $x$, defined as
$x = \left( \beta_b /\beta_b^* - 1 \right) L^{1/\nu}$
for a second-order transition, and
$x = L \exp\!\left[-\zeta \, \big|\beta_b / \beta_b^* - 1\big|^{-1}\right]$
for a Berezinskii–Kosterlitz–Thouless (BKT) transition, assuming the renormalization-group $\beta$ function vanishes quadratically. First-order transitions are also described by the first scaling form but with $\nu = 1/d = 0.25$ in our system.

\begin{figure}[tb]
    \centering
    \includegraphics[width=0.49\linewidth]{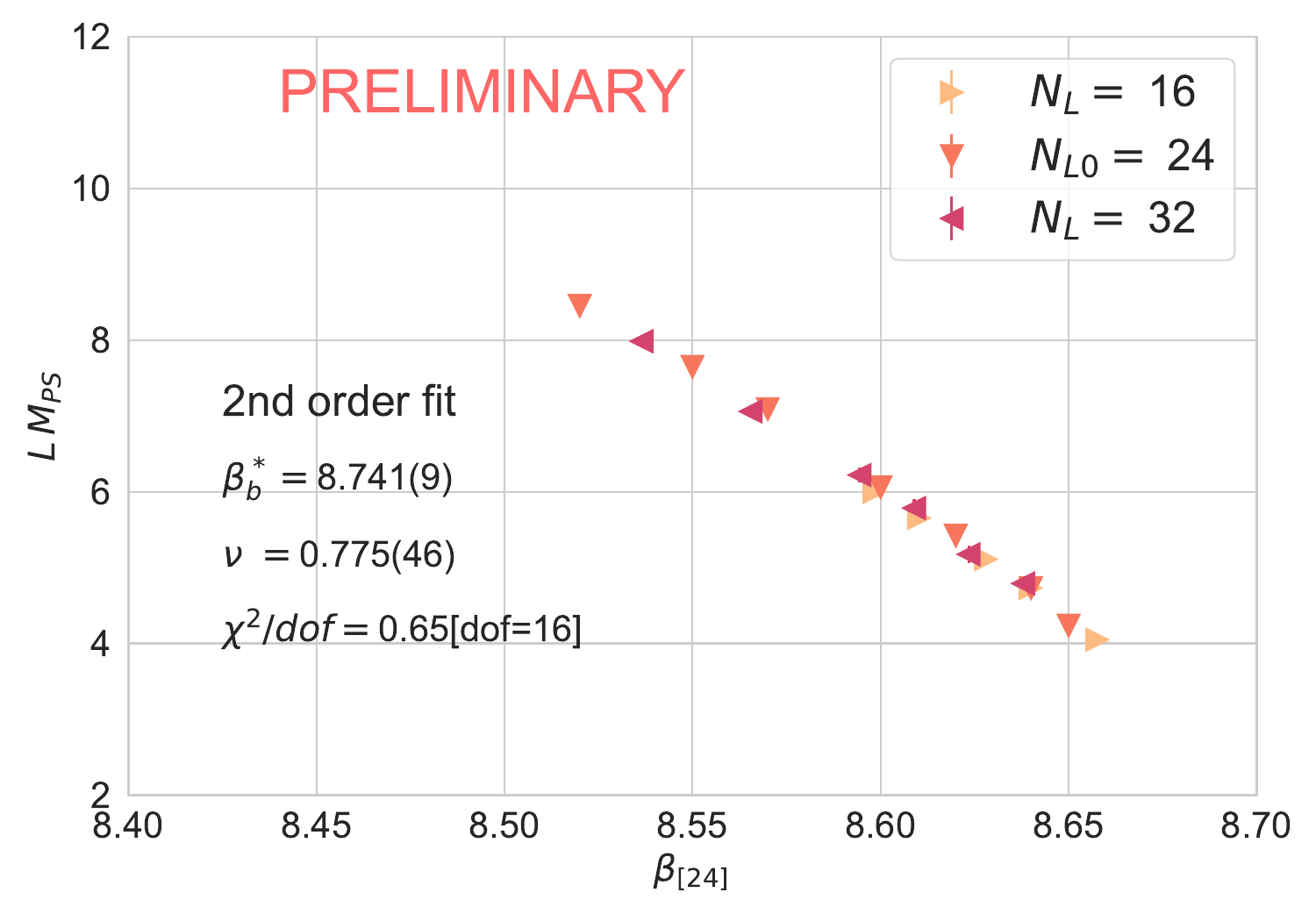}
     \includegraphics[width=0.49\linewidth]{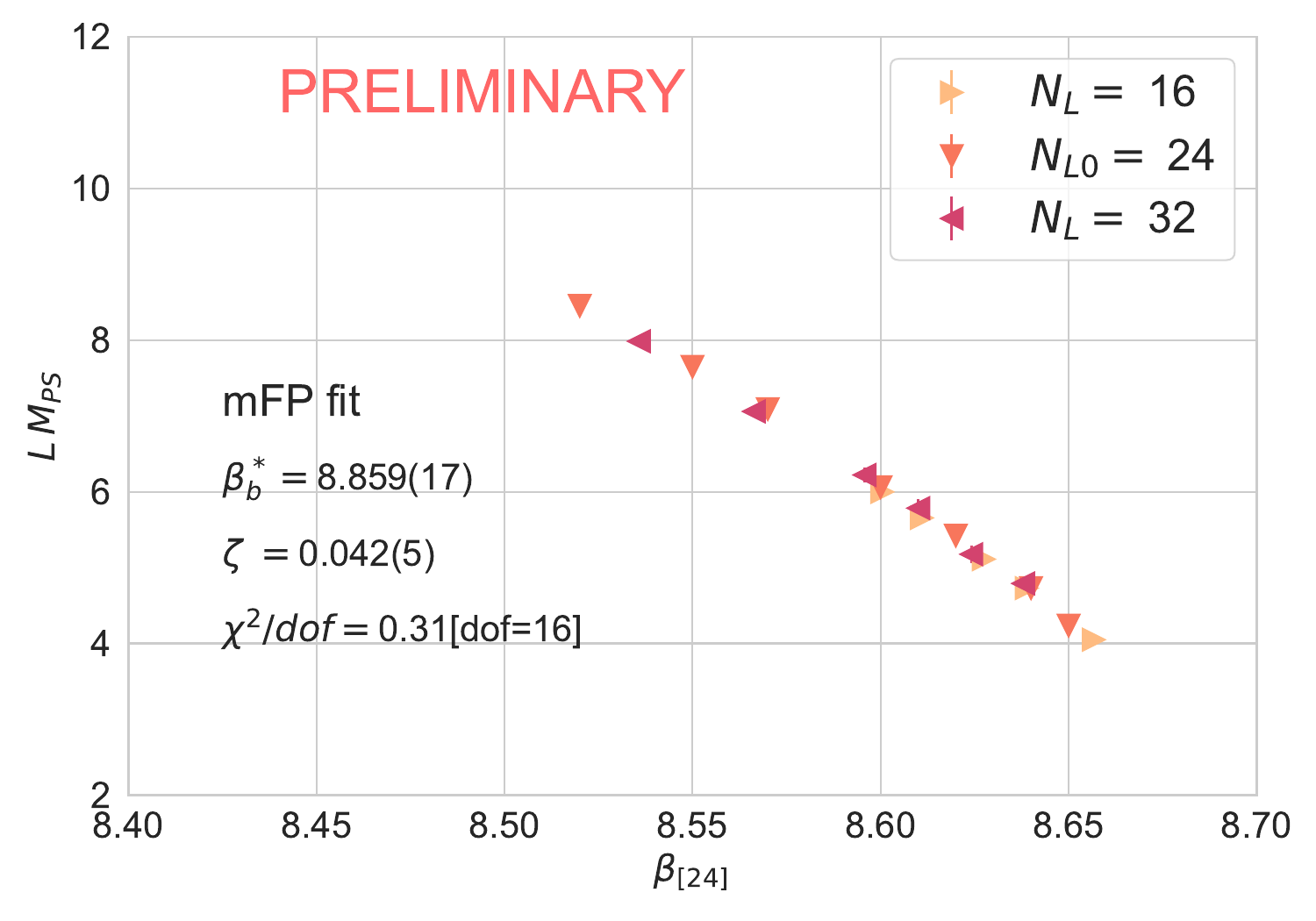}
    \caption{Curve-collapse fits to explore the nature of the phase transition using data near the critical point. The left plot tests the hypotheses of a 1\textsuperscript{st} or 2\textsuperscript{nd} order phase transition, whereas the plot on the right tests for a BKT transition.  }
    \label{Fig.M_PS-curvecollapse}
\end{figure}

As shown in Fig.~\ref{Fig.M_PS-curvecollapse}, we are not yet able to unambiguously determine the nature of the transition, as both curve-collapse fits yield acceptable $\chi^2/\text{d.o.f.}$ However, the fit testing the second-order scenario gives an exponent $\nu = 0.775(46)$, several standard deviations away from $1/d = 0.25$, the value expected for a first-order transition. We therefore rule out a first-order phase transition and conclude that the transition is continuous, defining a UV-complete quantum field theory. To further substantiate these findings, simulations on larger $48^3 \times 96$ volumes are highly desirable.

\section{Summary}

We have presented our investigation of the $SU(3)$ gauge theory with eight fundamental flavors at strong coupling, using nHYP-smeared staggered fermions with the Symanzik gauge action and additional Pauli–Villars fields. Our results indicate that this system exhibits a weak-coupling phase with conformal hyperscaling and a strong-coupling phase characterized by a gapped, chirally symmetric spectrum --- consistent with the presence of an SMG phase. The two phases are separated by a phase transition that appears to be continuous, possibly of BKT type; a first-order transition is excluded by our finite-size scaling analysis.

If confirmed, these findings imply that $SU(3)$ gauge theory with $N_f = 8$ flavors is conformal. More importantly, the existence of a continuous transition suggests the presence of an ultraviolet fixed point where a UV-complete continuum quantum field theory can be defined. Based on the symmetry structure of staggered lattice fermions, this continuum theory must include a four-fermion or other higher-dimensional operator that is irrelevant at the perturbative fixed point but becomes relevant at the SMG transition.

Further investigation of the infrared properties of the SMG phase is essential, particularly regarding Lorentz invariance and potential flavor-symmetry breaking. On the weak-coupling side, additional studies are needed to determine whether the system possesses a separate infrared conformal fixed point or whether the RG $\beta$ function vanishes quadratically at the new fixed point, signaling the opening of the conformal window.

\acknowledgments
AH.~acknowledges support from DOE grant DE-SC0010005.

The numerical simulations were performed using the Quantum EXpressions (QEX) code  \cite{Osborn:2017aci,Jin:2016ioq} for gauge field generation, Qlua \cite{Pochinsky:2008zz,qlua} for gradient flow measurements and hadronic spectra are calculated using MILC \cite{MILC}. We thank in particular James Osborn, Xiaoyong Jin, and Curtis T. Peterson for their assistance in starting this project.

Computational resources provided by the USQCD Collaboration, funded by the Office of Science of the U.S.~Department of Energy, Lawrence Livermore National Laboratory (LLNL), Boston University (BU) computers at the MGHPCC, in part funded by the National Science Foundation (award No.~OCI-1229059), and the OMNI cluster of the University of Siegen were used. We thank the LLNL Multiprogrammatic and Institutional Computing program for Grand Challenge supercomputing allocations.

This document was prepared using the resources of the USQCD Collaboration at
the Fermi National Accelerator Laboratory (Fermilab), a U.S.~Department of
Energy (DOE), Office of Science, Office of High Energy Physics HEP User Facility. Fermilab is managed by Fermi Forward Discovery Group, LLC, acting under Contract No.~89243024CSC000002.
{
  \bibliography{newBSM.bib}
  \bibliographystyle{JHEP}
}


\end{document}